# Hybrid gold single crystals incorporating amino acids

*Linfeng Chen,[1] Iryna Polishchuk,[1] Eva Weber,[1] Andy N. Fitch,[2] Boaz Pokroy[1]\**

[1]Department of Materials Science and Engineering and the Russel Berrie

Nanotechnology Institute, Technion-Israel Institute of Technology 32000, Haifa,

Israel.

[2]European Synchrotron Radiation Facility, B. P. 220, 38043 Grenoble Cedex,

France.

ABSTRACT: Composite hybrid gold crystals are of profound interest in various research areas ranging from materials science to biology. Their importance is due to their unique properties and potential implementation, for example in sensing or in bio-nanomedicine. Here we report on the formation of hybrid organic-metal composites via the incorporation of selected amino acids (histidine, aspartic acid, serine, glutamine, alanine, cysteine, and selenocystine) into the crystal lattice of single crystals of gold. We used electron microscopy, chemical analysis and high-resolution synchrotron powder X-ray diffraction to examine these composites.

Crystal shape, as well as atomic concentrations of occluded amino acids and their impact on the crystal structure of gold, were determined. Concentration of the incorporated amino acid was highest for cysteine, followed by serine and aspartic acid. Our results indicate that the incorporation process probably occurs through a complex interaction of their individual functional groups with gold atoms. Although various organic−gold composites have been prepared, to the best of our knowledge this is the first reported finding of incorporation of organic molecules within the gold lattice. We present a versatile strategy for fabricating crystalline nanohybrid-composite gold crystals of potential importance for a wide range of applications.

INTRODUCTION

Bio-inspired and biogenic composites have attracted increasing attention over the past few decades owing to their fascinating physical and chemical properties. They display mechanical,[1] optical,[2] magnetic,[3] electronic,[4] and wetting properties,[5] some of which have promising potential applications in medicine,[6] environmental systems,[7] optics,[8] sensing and detection,[9] and self-cleaning.[10] Ability to control the various components of composite materials is a key issue in regulating their overall properties. In this respect, organisms that deposit biominerals via the process of biomineralization provide a source of inspiration for fabricating nanocomposites. The exceptional mechanical properties of various biominerals derive from the interaction of their organic and inorganic phases and from their hierarchical organization,[11] as in sea urchin exoskeleton, mollusk shells, or bone. As little as 1 wt. % of intracrystalline macromolecules can significantly enhance the fracture toughness of the host brittle crystal.[12]

In recent years, numerous studies have shed light on the structure and microstructure of biominerals. Various techniques, such as high-resolution synchrotron powder X-ray diffraction (HRPXRD) and Transmission Electron Microscopy (TEM) have shown that organic macromolecules are indeed incorporated into individual crystallites,[13] significantly altering their lattice parameters and microstructure.[14] More recent studies have described the incorporation of single amino acids into individual synthetic crystals, leading to lattice distortions comparable to those found previously in biogenic samples.[15,16] This bioinspired strategy has been extensively applied to produce artificial materials with enhanced mechanical, optical, and electronic properties.[15a, 17] Besides amino acids, moreover, polymer micelles,[17b] polymer beads,[18] gels,[19,20] fullerenol nanoparticles,[21] single-walled carbon nanotubes,[22] dyes[23] and even anti-cancer drugs[24] have been incorporated. In these latter cases the applied incorporation was mainly into calcium carbonate rather than into ZnO host materials. To the best of our knowledge, there are no reports of incorporation of organic molecules such as amino acids into *metallic* crystalline hosts, including gold or coinage metals. In a pioneering work by Prof. David Avnir on molecularly doped metals,[25] various small molecules were entrapped within a framework of nanocrystals of some metals, including gold; they did not, however, penetrate the dense metal lattice. As gold nanomaterials have shown great potential in medicine,[26] plasmonics,[27] detection,[28] electronics[29] and catalysis,[30] such a bioinspired approach based on incorporation into gold crystal hosts could be important both for basic research and for applied scientific purposes.

Here we describe the one-step synthesis of hybrid nanocomposite gold crystals via occlusion of amino acids within the gold host crystals. As in previous studies with calcium carbonate[15b] and ZnO,[31] we found that the lattice constant of the gold crystalline host increased as a result of the amino acid incorporation. After mild annealing the lattice distortions were effectively relaxed as a result of decomposition of the organic molecules. The observed incorporation correlated with the binding affinity between gold atoms and various functional chemical groups of specific amino acids. This work provides a versatile strategy for the design and fabrication of novel hybrid gold−single crystalline nanocomposites.

EXPERIMENTAL SECTION

**Materials.** Amino acids, L-serine (Ser, 98.5−101.0%), L-cysteine (Cys, ≥98.5%), L-glutamine (Gln, 99.0−101%), L-histidine (His, ≥99%), L-alanine (Ala, ≥98.5%), and L-aspartic acid (Asp, ≥98%), as well as polyvinylpyrrolidone (PVP-40) were purchased from Sigma. Seleno-L-cystine (Sec, 98%) was purchased from ACROS Organics. Hydrated chloroauric acid containing Au 49% (99.99% pure) was purchased from Alfa Aesar. Ethylene glycol (≥99.5%) was purchased from Merck.

**Instruments and characterizations.** For high resolution scanning electron microscopy (HRSEM) we used a Zeiss Ultra-Plus FEG-SEM. For wavelength-dispersive X-ray spectroscopy (WDS) we used an FEI E-SEM Quanta 200. The working voltage was 10 kV. AlN was used to calibrate the background signal. At least 10 points were collected for detecting each sample. For transmission electron microscopy (TEM) we used an FEI Titan 80−300 KeV S/TEM.

**Synthesis of gold particles.** Gold crystals were synthesized by the polyol reduction method as described.[32] Briefly, ethylene glycol (20 mL) and PVP (340 mg) were mixed in a clean beaker with a stopper, and this was followed by magnetic stirring for about half an hour. $HAuCl_4$ (250 mM, 0.495 mL) was then added without stirring to the solution, which was heated in an oven at 120°C for 20 h. As a control, gold crystals were obtained by repeated centrifugation and washing with deionized water and ethanol. Gold crystals with incorporated amino acids were prepared as for the control, with the addition of amino acids into the mixture during the first step while stirring. The final concentration for His, Asp, Ala, Ser, Cys, or Gln was 1 mg mL$^{-1}$. The final concentration for Sec**,** owing to its low solubility in water was 0.01 mg mL$^{-1}$.[33]

**Characterization of gold crystals by high-resolution powder X-ray diffraction.** HRPXRD measurements of gold crystal powders were performed at the ESRF on a dedicated ID22 beamline at a wavelength of 0.490069Å (±0.000004Å). The beamline is equipped with a crystal monochromator and a crystal-analyzer optical element at the incident and diffracted beams, respectively, yielding diffraction spectra of superior quality and, importantly, intense and extremely narrow diffraction peaks with an instrumental contribution to the peak widths not exceeding 0.004 degrees.[34] Powders were loaded into 0.7-1 mm borosilicate glass capillaries. Lattice parameters were extracted utilizing the Rietveld refinement method with GSAS software and the EXPGUI interface.[35]

RESULTS AND DISCUSSION

Inspired by reports of incorporation of amino acids into various crystalline hosts, we investigated the feasibility of using the same strategy for their incorporation into pure metallic single crystals. For this first study we chose gold. Hybrid composite crystals in the presence of amino acids were prepared from their corresponding solutions according to the polyvinylpyrrolidone (PVP)-mediated polyol-reduction method, as described in Methods and in previous reports.[27, 36,32] To study the possible impact of amino acids on the crystal structure of gold and their potential for incorporation, we selected seven amino acids with different functional groups: Histidine (His) − positively charged side chain; aspartic acid (Asp) − negatively charged side chain; serine (Ser) − hydroxyl residue in the side chain; glutamine (Gln) − amine residue in the side chain; alanine (Ala) − non-polar side chain; cysteine (Cys) − thiol group in the side chain; and selenocystine (Sec) − selenol group in the side chain. Each amino acid was introduced into its reaction solution at a final concentration of 1 mg mL$^{-1}$ with the exception of Sec (0.01 mg mL$^{-1}$) because of its relatively low solubility in water. In the rest of this paper the gold−amino acid (Au-AA) composites are abbreviated as Au-His, Au-Asp, Au-Ser, Au-Gln, Au-Ala, Au-Cys, and Au-Sec.

As a first step we investigated the morphology of crystals grown in the presence of the various amino acids by using high-resolution scanning electron microscopy (HRSEM). We observed that gold crystals in the absence of amino acids (Au-control) appeared as a mixture of various shapes, such as triangular plates, truncated triangular plates and pentagonal bipyramids, with sizes ranging from 300 nm to 1 μm (Figure 1a). This observation is similar to those reported elsewhere.[32, 37] Where amino acids were present, by contrast, the shapes and sizes of the gold crystals were significantly affected (Figure 1b—h). With the exception of Au-Sec, all crystals grown in the presence of an amino acid were reduced in size. The predominant morphology of Au-His is an icosahedron with a wide size range of 100−400 nm (Figure 1b). The most intriguing change in shape, however, with flat crystals greater than 10 μm, was observed for Au-Sec (Figure 1h).

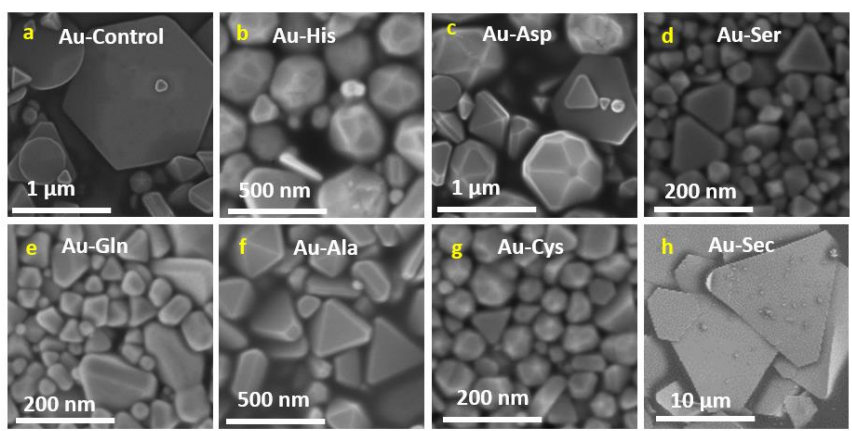

**Figure 1.** HRSEM images of Au crystals prepared by the PVP-mediated polyol-reduction process in the absence of amino acid (a, Au-control) and in the presence of histidine (b, Au-His), aspartic acid (c, Au-Asp), serine (d, Au-Ser), glutamine (e, Au-Gln), alanine (f, Au-Ala), or cysteine (g, Au-Cys) at a concentration of 1 mg mL$^{-1}$, and selenocystine (h, Au-Sec) at a concentration of 0.01 mg mL$^{-1}$.

The nano and atomic structures of the crystals grown in the presence of an amino acid were further characterized by means of aberration-corrected transmission electron microscopy (TEM, Figure S1). Our findings indicated that individual Au crystals grown in the presence of amino acids are indeed single crystalline, as shown by fast Fourier transforms (FFT) applied to lattice images (Figure S1).

To find out whether the amino acids had become incorporated into the single crystals of gold, we first analyzed the Au crystals chemically. We were unable to use the technique that we had previously utilized for calcite[38], namely dissolution of crystals followed by amino acid analysis because the known methods for dissolving gold are harsh and thus destroy the amino acids. We therefore chose wavelength-dispersive X-ray spectroscopy (WDS) as our method of chemical analysis. All experiments were performed only after surface-bound organic molecules had been removed from the samples by sodium borohydride (NaBH4) treatment coupled to sonication.[39] The atomic concentration of each amino acid was calculated on the basis of its nitrogen concentration (Table 1). At the 95% confidence level, considerably low nitrogen signal was detected from the Au-control and much lower than its detection limit, indicating that the incorporation of PVP, which also contains nitrogen, was negligible. The existence of nitrogen was confirmed for all other samples except the case of Au-Sec with signal lower than a detection limit of 0.2 at.%. This exception might be explained by the low concentration of Sec used in the experimental setup owing to its limited solubility in water.[33] The results showed that the atomic concentration of amino acid was highest in Au-Cys and the lowest in Au-His (Table 1).

**Table 1.** Atomic concentration of the various amino acids (at.%) within Au crystals detected by WDS (p = 0.05).

| Sample | Amino acid (at.%) | Detection limit (at.%) |
|--------|-------------------|------------------------|
| Au-control | 0.07±0.06 | 1.38 |
| Au-Cys | 1.56±0.43 | 0.85 |
| Au-Asp | 0.79±0.23 | 0.44 |
| Au-Ser | 0.71±0.27 | 0.27 |
| Au-Ala | 0.57±0.10 | 0.21 |
| Au-Gln | 0.40±0.09 | 0.18 |
| Au-His | 0.23±0.08 | 0.15 |
| Au-Sec | 0.19±0.10 | 0.20 |

In addition to the WDS experiments we also performed thermogravimetric analysis coupled with mass spectroscopy (TGA-MS). To this end we analyzed the Au-control as well as newly grown Au-Ala in the presence of 1 mg mL$^{-1}$ of alanine. Results of the TGA-MS analysis are presented in Figure 2.

Clearly, the hybrid Au-Ala lost much more weight than the control sample (0.35 ± 0.1 wt.% as compared to 0.1 wt.%). Furthermore, most of the weight loss occurred around 250ºC, just after MS emitted signals indicating chemical species with molecular masses of 18, 28 and 44 m/z. A mass of 18 probably corresponds to water, while a mass of 44 probably corresponds to $CO_2$ or $N_2O$, and a mass of 28 to $N_2$. Decomposition of amino acids has been previously shown to lead to emittance of $CO_2$.[40] The control sample demonstrated a very small water peak upon heating, probably owing to surface-bound humidity, and a small $CO_2$ peak that might reflect the decomposition of PVP. To confirm these results we performed WDS on the same Au-Ala sample as that used in the TGA-MS analysis. Ala was detected at a concentration of 0.77 ± 0.17 at%; this is equivalent to 0.35 ± 0.08 wt%, which fits the TGA results very well. However, the Ala concentration incorporated in gold crystals observed in this WDS experiment slightly differed from that seen in the first series of experiments, in which the Ala concentration detected was only $0.57 \pm 0.10$ at.% (see Table 1). This suggested that the incorporation is probably dependent on several parameters, such as growth rate, which were not highly controlled the earlier (proof-of-principle) experiment.

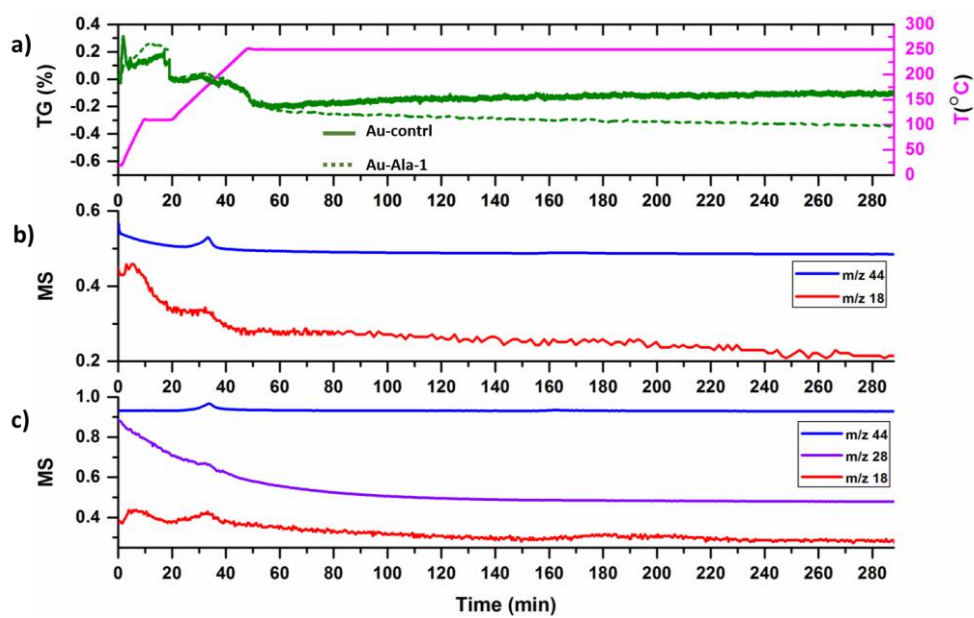

Figure 2: TGA-MS of Au-control and Au-Ala samples. a), TGA of the Au-control and Au-Ala samples. b) and c), MS analysis of the Au-control and Au-Ala samples respectively. Mass-to-charge ratios (m/z) of 18, 28 and 44 probably correspond to $N_2$, $H_2O$ and $CO_2/N_2O$ accordingly.

To further study the amino acid incorporation, we used HRPXRD to compare the characteristics of the crystal structure and microstructure of the Au-AA hybrids to that of pure Au. All experiments were performed with a dedicated high-resolution powder diffraction synchrotron beamline (ID22 at the European Synchrotron Radiation Facility (ESRF), Grenoble, France) and confirmed that all samples exhibited the gold structure (Figure S2). The most intense {111} diffraction peaks of the Au-control sample and the Au-Ala sample before and after annealing at 250ºC for 2 h are shown in Figure 3a and 3b, respectively. Comprehensive investigation of individual diffraction peaks nevertheless revealed that the diffraction peaks shifted to a smaller diffraction angle only for those samples grown in the presence of amino acids. This lowering of the diffraction angle corresponds to crystal lattice expansion. Moreover, after mild annealing the diffraction peaks reverted to the position of the Au-control, pointing to relaxation of lattice distortions after the amino acids are thermally affected. This result was consistent with our previous finding in carbonate systems, namely that heating leads to relaxation of the lattice distortions induced by occluded organic molecules.[14a, 15b] In contrast, the Au-control showed no change in peak position before and after heat treatment (Figure 3a). To confirm that the lattice expansion is not due to the crystal size effect we refer to Solliard et al., who showed that nanometric size of gold particles leads to lattice contraction and that this effect is prominent only below 40 nm.[41] That finding rules out such a potential cause and further confirms that amino acids indeed became incorporated into the single crystals of Au.

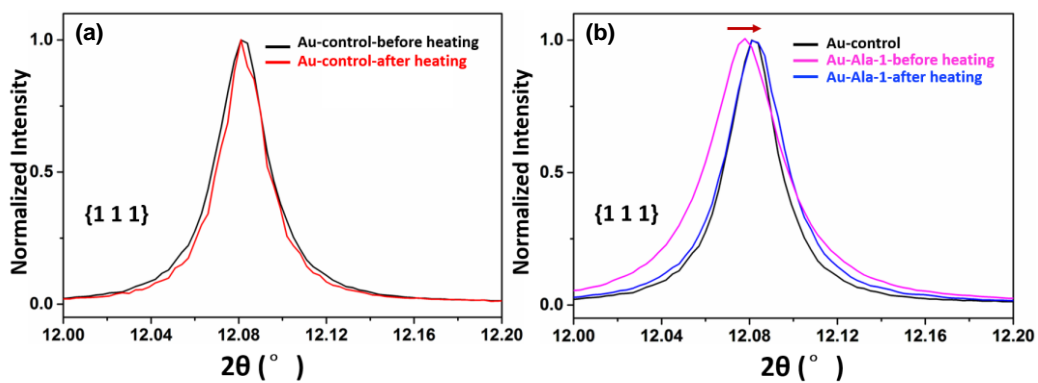

**Figure 3.** HRPXRD gold {111} diffraction peaks from Au-control and Au-Ala samples before and after heating. a) Au-control before (black) and after (red) annealing at 250ºC for 2 h. b) Au-Ala before (magenta) and after (blue) annealing at 250ºC for 2 h compared to Au-control (black) before annealing. Note that after annealing the diffraction peak of Au-His shifts back to the Au-control position (arrow).

To quantify the observed lattice distortions and correlate them with our chemical analysis data, we performed full-profile Rietveld refinement[42] and extracted the lattice parameters. A typical Rietveld refinement profile is shown in Figure S3. The structural parameters were obtained with the highest accuracy (ca. 10 ppm) and the lattice distortions were calculated according to the following formulation:

$$\frac{\Delta a}{a_a} = \frac{a_b - a_a}{a_a}$$

where $a_b$, $a_a$ represents the lattice constant before and after heating, respectively. The lattice distortions for all of the composite gold crystals are plotted in Figure 4 (absolute lattice parameters are presented numerically in Table S1). It can be seen that Au-Cys indeed exhibited the largest lattice distortions with a maximum value of $1.3 \cdot 10^{-3}$, followed by Au-His ($3.7 \cdot 10^{-4}$), Au-Asp ($3.1 \cdot 10^{-4}$), Au-Gln ($3.0 \cdot 10^{-4}$), Au-Sec ($2.5 \cdot 10^{-4}$), Au-Ala ($1.6 \cdot 10^{-4}$) and Au-Ser ($1.4 \cdot 10^{-4}$). While all samples exhibit a relative increase of lattice parameter, the lattice parameter of the control sample decreased very slightly upon annealing which might be attributed to relaxation of stresses caused by various defects generated during the metal crystal growth. Nevertheless, these lattice distortions were negligible in comparison to the lattice distortions caused by amino acid incorporation.

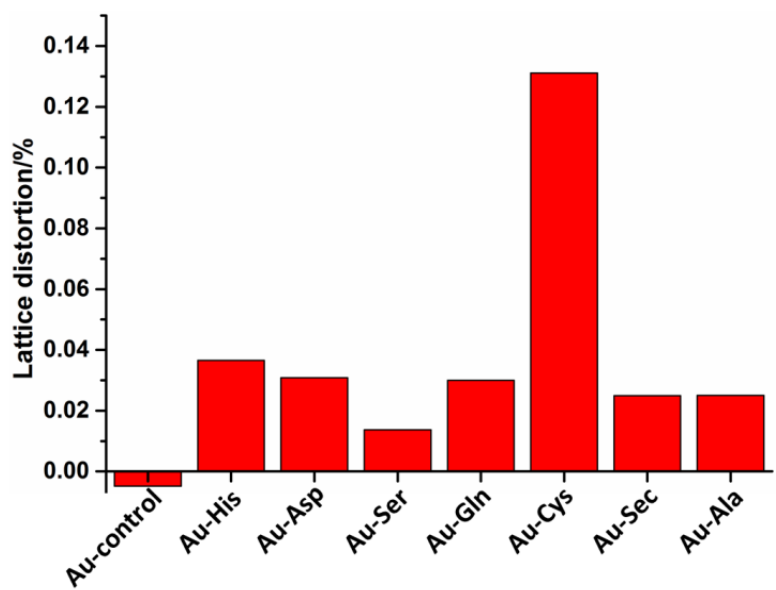

**Figure 4.** : Lattice distortions (relative to annealed samples).

We performed line profile analysis to obtain microstructural parameters of coherence length and micro-strain fluctuations. This was done as described in our previous paper.[14a] The derived values are presented in Table S2 and Figure S4. The table shows that the largest coherence length (the average distance between defects, not the physical crystal size) was highest for the control sample and was about 120 nm. As amino acids became incorporated the average coherence length decreased slightly for most of the hybrid crystals, as expected. The coherence length was lowest for incorporation of Cys, which also demonstrated the highest lattice distortions. After annealing, all samples exhibited diffraction peak narrowing (see Figure 3) owing to the increase in coherence length and decrease in microstrain fluctuations. This was in contrast to our observations for biogenic crystals[14a] and for amino acid incorporation into synthetic calcite[15b]. The difference can probably be explained by the much higher diffusion rates of gold than of carbonates at these lower temperatures, allowing for crystal growth and lowering the defect concentration in gold.

The finding that the highest concentration of incorporation, leading in turn to the highest lattice distortions, was shown by Cys might be explained by the fact that Cys exhibits a high affinity for gold owing to the strong interaction between the thiol group and Au (the energy of the thiolate-gold bond energy is ca. 40 kcal mol$^{-1}$).[43] As for the other amino acids, we believe that their incorporation can be induced by the well-known affinity of amino groups for gold,[44] albeit weaker than the affinity of thiolate groups. It has been shown that such additives often become incorporated via step edges on the growing crystal.[45]

To study the relative lattice distortions induced by the various amino acids, we normalized the lattice distortions to amino acids concentrations, as determined by WDS (Figure 5). Interestingly, the highest normalized lattice distortions, though detected in rather low concentrations, were induced by His (Table1).

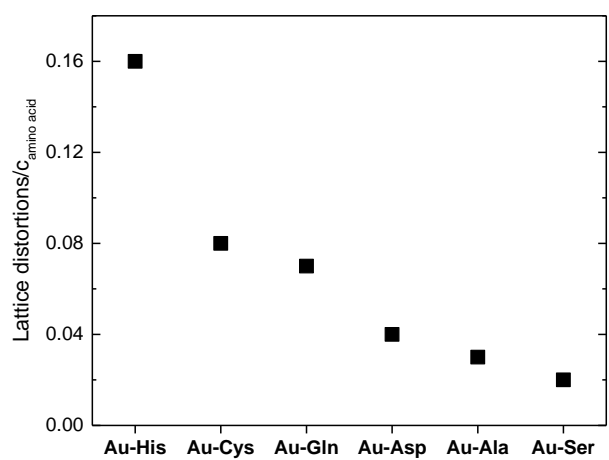

**Figure 5.** Lattice distortions normalized to amino acid concentrations, as determined by WDS, demonstrating the ability of amino acids to induce lattice distortions.

On the other hand, Ser, Asp and Ala, despite being detected at higher concentrations, induced lower normalized lattice distortions. Although the mechanism of incorporation is not yet understood, it seems reasonable to suggest that the differences in relative lattice distortions derive from the composition of the amino acids and their 3D organization within the gold crystal. This would indicate that His, Cys and Gln (in which the normalized lattice distortions are higher) do not fit well in the lattice. His contains a rigid imidazole functional group, Cys encloses a thiol active site, whereas Gln, in addition to its amino group, has an amine functional group. The latter might also bind to gold and therefore permit less reorganization within the lattice.

Better understanding of the incorporation mechanism and the organization of each amino acid within the gold lattice will require systematic computational modeling. This, however, is a topic for a separate study.

CONCLUSIONS

We showed in this study that amino acids can become incorporated into the crystal lattice of single crystals of gold. To the best of our knowledge, this is the first reported example of the incorporation of an organic species within the lattice of a metal single crystal. We found that various amino acids become incorporated and that Cys, which is known to bind strongly to gold via the thiol group, becomes incorporated at the highest concentration. We believe that the other amino acids also interact with gold via their amine groups.

The ability to synthesize such organic-metal hybrid single crystals is likely to have potential applications in nano-biomedicine, in which gold particles are utilized in various sensing devices, and in other applications. Further investigation, including computational modeling and various functional studies, are called for and will be the focus of our next steps.

ASSOCIATED CONTENT

**Supporting Information**. TEM results, XRD pattern of gold, absolute values of lattice parameter for every sample, Rietveld refinement plot, microstructure parameters before and after heating. This material is available free of charge via the Internet at http://pubs.acs.org.

AUTHOR INFORMATION

**Corresponding Author**

*E-mail: bpokroy@tx.technion.ac.il

ACKNOWLEDGMENTS

The research leading to these results received funding from the European Research Council under the European Union's Seventh Framework Program (FP/2007-2013)/ERC Grant Agreement (no. 336077). L.C. acknowledges financial support by a fellowship from the Israel Council for Higher Education and the Technion Fund for Cooperation with Far Eastern Universities. We thank Dr. Alex Berner for help in collecting the WDS data and Dr. Davide Levy for useful discussions on the Rietveld refinement results.